\documentclass[fleqn,10pt]{SelfArx} 

\usepackage[english]{babel} 

\usepackage{lipsum} 

\definecolor{color1}{RGB}{0,0,90} 
\definecolor{color2}{RGB}{0,20,20} 

\usepackage{hyperref} 
\usepackage{xcolor}
\usepackage{graphicx}
\usepackage{epstopdf, epsfig}
\usepackage{mathtools}  
\usepackage{amsmath}
\hypersetup{
	hidelinks,
	colorlinks,
	breaklinks=true,
	urlcolor=color2,
	citecolor=color1,
	linkcolor=color1,
	bookmarksopen=false,
	pdftitle={Title},
	pdfauthor={Author},
}

\JournalInfo{Research Article} 
\Archive{} 

\PaperTitle{On the Thermofluidics of a Steady Laminar Jet Impacting on a Rotating Hot Plate: An ab-initio Scaling Perspective} 

\Authors{Durbar Roy\textsuperscript{1} and Saptarshi Basu\textsuperscript{1}*} 
\affiliation{\textsuperscript{1}\textit{Department of Mechanical Engineering, Indian Institute of Science, Bengaluru, KA 560012, India}} 

\affiliation{*\textbf{Corresponding author}: sbasu@iisc.ac.in} 


\newcommand{\keywordname}{Keywords} 

\Abstract{We provide an ab-initio scaling analysis for liquid film thickness and Nusselt number of a steady laminar jet impacting a rotating heated plate. The limiting scaling regimes incorporating the evaporative effects along the liquid-vapor interface have been probed. The dependence of liquid film thickness and Nusselt number on Reynolds, Rossby, and Prandtl number have been unearthed using scaling analysis of the integral and differential form of the continuity, Navier-Stokes and energy equation in a cylindrical coordinate system. Boundary layer analysis has been used to discover a critical length that plays a significant role in understanding the effect of evaporation on hydrodynamic, thermal boundary layer thicknesses, and subsequently Nusselt number. The evaporative effects on liquid film thickness become increasingly important after a certain critical radius. The scaling laws derived were compared with existing experimental data available in the literature, and the trends predicted were consistent.}
\begin{document}
\maketitle
\section{Introduction}
Many scientists and engineers have pursued the hydrodynamics and heat transfer of liquid jet impacting on a rotating heated plate in the past \cite{rice2005analysis,webb1995single,ozar2004experiments}. Liquid jets find much interest among the community due to their applications in various heat transfer devices like radiators, evaporators in aerospace applications where efficiency is of utmost importance. Jets are considered effective in transferring heat and mass due to thin hydrodynamic and thermal boundary layer effects. A lot of literature available considers a variety of conditions like the effect of rotation, wall temperature, surface tension, to name a few \cite{thomas1990one} on heat transfer dynamics. 
Researchers have also analyzed liquid film thickness, Nusselt number and hydraulic jump using experimental, numerical, semi-analytical and analytical techniques \cite{watson1964radial}. Turbulent flow computations using $k-{\epsilon}$ model in thin liquid fluid layers involving a hydraulic jump were accomplished by a group led by Rahman \cite{rahman1991computation}. Rahman et al. worked out the complete numerical solutions of the momentum equations using finite difference techniques incorporating a boundary-fitted coordinate technique. Rahman and Faghri also carried out detailed works on rotating plate using a three-dimensional boundary-fitted coordinate system \cite{rahman1992numerical}. They concluded the dominance of inertia at the entrance of the jet and rotation at the outer edge of the plate. Micro-gravity experiments and numerical simulations using potential flow theory were performed for free circular jets impinging on flat plates by scientists at NASA \cite{labus1977liquid}. Avedisan and co-workers also studied circular hydraulic jumps in low gravity environments \cite{avedisian2000circular}. Scientists studied free liquid jets using semi-analytical techniques like the work of Rao and co-workers \cite{rao1998integral}. Rao et al. used an integral analysis for the boundary layer equations using third-order polynomials approximations for the velocity profiles. Heat transfer characteristics were calculated using similar integral analysis like Rao and co-workers by Liu and Lienhard \cite{liu1989liquid}. They performed the computations for a uniform heated surface without rotations and discussed the effect of Prandtl number on Nusselt number. The heat transfer analysis for jets impacting on rotating plates were studied by Basu and Cetegen \cite{basu2006analysis}. They used integral analysis techniques to calculate liquid film thickness and Nusselt number.
Azuma and Hoshino studied the transition from laminar to turbulent, performed stability analysis, calculated liquid film thickness, and wall pressure fluctuations for stationary horizontal plate \cite{azuma1984radial}. The heat transfer characteristics, including Nusselt number calculations, were first studied by Chaudhury and group \cite{chaudhury1964heat}. They developed closed-form solutions in the region where similarity solution exists. The heat transfer at the boundary between the liquid film and solid plate was solved separately, and then the solutions were matched at the boundary by Wang and his colleagues \cite{wang1989heat}. The convective heat transfer from a jet of cooling oil to an approximately isothermal rotating plate was calculated by Carper and co-workers \cite{carper1978heat}. A later study by the same group incorporated Prandtl number in the heat transfer characteristics. They provided many correlations from experimental data fitting. The numerous correlations developed in literature
\cite{carper1986liquid} lacked fundamental physical and mathematical insights from first principles. 
However, scientific literature, that includes the evaporative effects on liquid film thickness and Nusselt Number in the case of a laminar free jet impinging on a rotating heated plate, is relatively sparse.
\\\\
We investigate liquid film thickness and Nusselt number in the high Froude number limit or equivalently in near zero-gravity conditions for a laminar axisymmetric free jet impinging on a rotating heated plate including evaporative effects.
Scaling analysis from first principles has been used to decipher the various limiting scales and the dependence of film thickness, the hydrodynamic and thermal boundary layers, Nusselt number on the geometric parameters, and various non-dimensional parameters such as Prandtl, Reynolds, and Rossby Number.
The complexity of the physics results firstly from the evaporative effects existing at the liquid-vapor interface compared to standard hydrodynamical interactions within a single liquid phase that are taken care by terms in the Navier-Stokes equation. Secondly, a coupling of the evaporative effects with the associated thermal energy equation exists, making the physics significantly more intricate.
We tackle the first complexity and incorporate evaporative effects using an integral mass conservation equation coupled with the standard differential form of mass and linear momentum equation.
The complexity of coupling the evaporative effects and the thermal energy equation is solved by incorporating the energy equation's thermal boundary layer analysis and the integral energy balance equation. The scalings of the fundamental equations were analyzed assuming constant evaporative flux. However, the results derived are general enough and can be extended to incorporate more complicated evaporative flux fields, which are not part of this study.
\begin{figure*}
  \centerline{\includegraphics[scale=0.4]{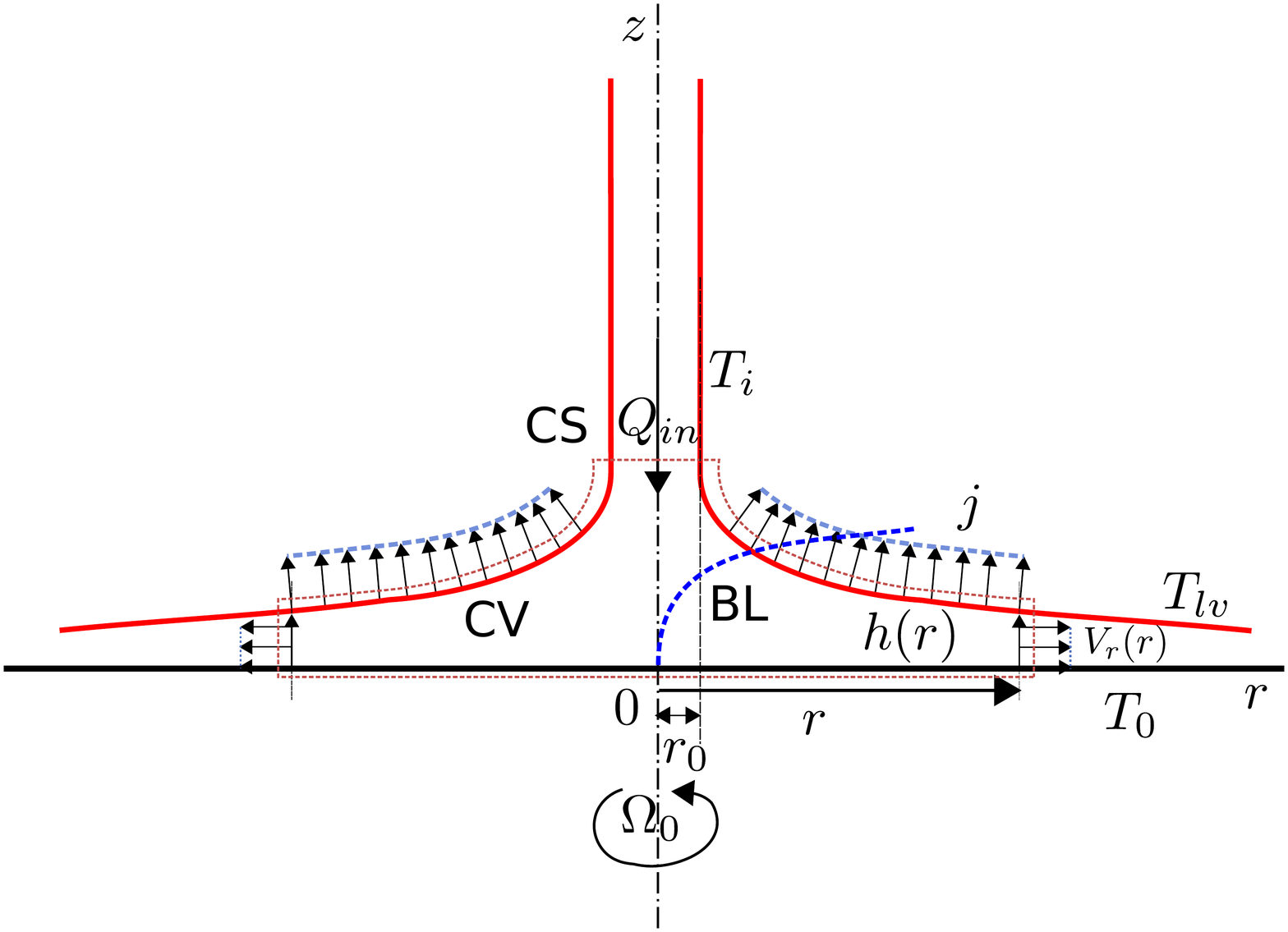}}
  \caption{Schematic representation of the axisymmetric laminar jet. $Q_{in}$ represents the volume inflow rate. $h(r)$ is the film thickness. ${\delta}(r)$ represents the boundary layer growth. $r_v$ denotes the radial location the growing boundary layer meets the film thickness due to the jet. The plate is heated after $|r|>r_0$, and ${\omega}_0$ denotes the rotational velocity of the disc. CS denotes the control surface (shown in dotted red) bounding the control volume CV.
The growing hydrodynamic boundary layer is denoted by BL shown by the blue dotted line.
    $T_i$ represents the temperature of the jet entering the control volume. $T_0$ is the bottom plate surface temperature and $T_{lv}$ is the liquid vapor surface temperature. The evaporative flux is denoted by $j$.}
\label{figure1}
\end{figure*}
\section{Mathematical modelling}

\subsection{Geometry and Coordinate System}
Figure 1 shows the schematic representation of a steady laminar jet impinging on a rotating heated plate.
We have used a cylindrical axisymmetric coordinate system about the vertical $z$ axis to formulate the conservation equations (mass, linear momentum and thermal energy). The jet cross-section lies parallel to the $r$ and $z$ plane with the plate at $z=0$. The jet impinges along the negative $z$ direction (axial direction). The radial coordinate $r$ has a length scale $r_0$, and the axial coordinate has a length scale $h_0$.
A circular rotating plate of an enormous radius $R$ compared to $r_0$ lies in the horizontal plane ($R>>r_0$).
The plate rotates at a constant angular velocity of ${\Omega}_0$. The inflow volume flow rate of the laminar jet is $Q_{in}$. The liquid film thickness profile is denoted by $h(r)$. ${\delta}(r)$ is the hydrodynamic boundary layer thickness which grows with the radial coordinate $r$.  The radial distance from the origin where ${\delta}(r)$ reaches a maximum is defined as $r_v$.
The bottom plate is heated from $|r|{\geq}r_0$.

Conservation of mass, momentum, and energy in the cylindrical coordinate system has been used to develop the various limiting scaling laws. The evaporation from the liquid film is incorporated in the analysis through a non-dimensional evaporation flux parameter.

\subsection{Integral Mass conservation}
The jet, upon impinging the circular plate, rotating at an angular velocity of ${\Omega}_0$ has an initial transient period. On reaching a steady state, the liquid film thickness profile $h(r)$ does not change anymore with time.
Applying integral mass conservation inside the control volume (refer to figure 1) at steady state, we have \cite{batchelor2000introduction,kundu2008fluid}
    \begin{equation}
Q_{in}=V_r(r)2{\pi}rh(r)+\frac{1}{\rho}{\int_{r_0}^{r}2{\pi}r{\sqrt{1+{\left(\frac{{\partial}h}{{\partial}r}\right)}^2}}\mathbf{J}{\cdot}{\mathbf{n}}dr}
      \end{equation}
where $V_r(r)$ denotes the radial velocity averaged over the liquid film thickness.
      The first term on the left-hand side of equation (1) is the volume flow rate of the impacting jet entering the control volume. The second term on the right-hand side represents the volume of liquid flowing out of the control volume radially, and the third term on the right-hand side represents the mass loss from the control volume due to evaporative effects from the liquid film and air interface.
      Approximating the above equation by a scaling equivalent form by neglecting second order curvature effects and using $\mathbf{J}{\cdot}{\mathbf{n}}{\sim}const$, $\mathbf{J}{\cdot}{\mathbf{n}}{\sim}j=|D{\nabla}c|$ where $c$ is the concentration field at the water air interface.
 Equation (1) reduces to
      \begin{equation}
Q_{in}{\sim}V_r(r)2{\pi}rh(r)+{\frac{j{\pi}}{\rho}}(r^2-r_0^2)
        \end{equation}
      Rearranging and making average radial velocity the subject, $V_r(r)$ scales as
\begin{equation}
          V_r(r){\sim}\frac{1}{2{\pi}rh(r)}[Q_{in} - {\frac{j{\pi}}{\rho}}(r^2-r_0^2)]
        \end{equation}
        \subsection{Differential form of continuity equation}
        The differential form of continuity equation in an axisymmetric cylindrical coordinate system is \cite{batchelor2000introduction,kundu2008fluid}
        \begin{equation}
          \frac{1}{r}\frac{{\partial}(rV_r)}{{\partial}r}+\frac{{\partial}V_z}{{\partial}z}=0
        \end{equation}

The radial coordinate $r$ scales as $r_0$ and hence equation (4) scales as
   \begin{equation}
\frac{V_r}{r_0}{\sim}\frac{V_z}{z}
\end{equation}
Using the stream function formulation in cylindrical coordinates ${\psi}(r,z)$ for the radial and axial velocity field, we are assured that conservation of mass is valid at every point inside the control volume
          \begin{equation}
          V_r=-\frac{1}{r}\frac{{\partial}\psi}{{\partial}z}
          \end{equation}
          and
          \begin{equation}
          V_z=\frac{1}{r}\frac{{\partial}{\psi}}{{\partial}r}    
          \end{equation}
          where $V_r$ and $V_z$ denotes the radial and axial velocity components of the liquid film respectively.  
     \subsection{Differential stream function formulation of radial Momentum equation}
The stream function formulation \cite{batchelor2000introduction,kundu2008fluid} for the radial component of the Navier Stokes equation under the thin film limit is given by

\begin{equation}
  \frac{1}{r}\frac{\partial}{{\partial}r}\left[\frac{1}{r}{\left(\frac{{\partial}\psi}{{\partial}z}\right)}^2\right]-\frac{\partial}{{\partial}z}\left[\frac{1}{r^2}\frac{{\partial}\psi}{{\partial}z}\frac{{\partial}\psi}{{\partial}r}\right]={\Omega}_0^2r-{\nu}\frac{{\partial}^2}{{\partial}z^2}\left[\frac{1}{r}\frac{{\partial}\psi}{{\partial}z}\right]
  \end{equation}
 
 Introducing the scales of the various terms in equation (8).
 The average radial velocity scales as
 $V_r{\sim}V_*$,
the stream function scales as 
${\psi}{\sim}rV_*z$,
the axial velocity scales as
${V_z}{\sim}\frac{1}{r}V_*z$.
Therefore equation (8) can be rewritten in a scaling form as
\begin{equation}
\frac{V_*^2}{r}{\sim}{\Omega_0^2r},{\frac{{\nu}V_*}{z^2}}
\end{equation}
The first term on the left hand side of equation (9) represents the inertial term. The second term and the third term on the right hand side denotes the rotation and viscous effects respectively.
\subsection{Differential form of energy equation}
The differential form of the thermal energy equation \cite{bejan2013convection} for the temperature field $T(r,z)$ is 
 \begin{equation}
   \frac{1}{r}\frac{\partial}{{\partial}r}{\left(rV_rT\right)}+
   \frac{\partial}{{\partial}z}\left(V_zT\right)=
   {\alpha}\left(\frac{{\partial}^2T}{{\partial}z^2}+\frac{1}{r}\frac{\partial}{{\partial}r}\left(r\frac{{\partial}T}{{\partial}r}\right)\right)
 \end{equation}

Using the scales $r{\sim}r_0$, $z{\sim}{\delta}_T$ the scaling form of equation (10) in the thermal boundary layer limit becomes
\begin{equation}
  \frac{V_r{\Delta}T}{r_0},\frac{V_z{\Delta}T}{\delta_T}{\sim}\frac{{\alpha}{\Delta}T}{{\delta}_T^2}
\end{equation}
The first and second term on the left hand side represents the radial and axial convective terms respectively. The third term on the right hand side represents the axial thermal diffusive term.
The fundamental relations in this section like equation (3), (5), (9), (11) have been used to derive the scalings for liquid film thickness and Nusselt number as presented in the results and discussion section.

\section{Results and Discussions}
The analysis required to calculate the scales of liquid film thickness and Nusselt number depends on various limiting conditions discussed below in separate subsections. The scale of liquid film thickness height profile $h(r)$ and Nusselt number $Nu(r)$ is an intricate function that depends on the interaction between the inertial, rotation, viscous, and evaporative effects. We will start by deciphering the scales of the liquid film thickness under various limiting conditions.

\subsection{Liquid film thickness}
\subsubsection{Inertial Inviscid Rotation Scaling with variable scale for radial velocity}
\begin{figure}
  \centerline{\includegraphics[scale=0.7]{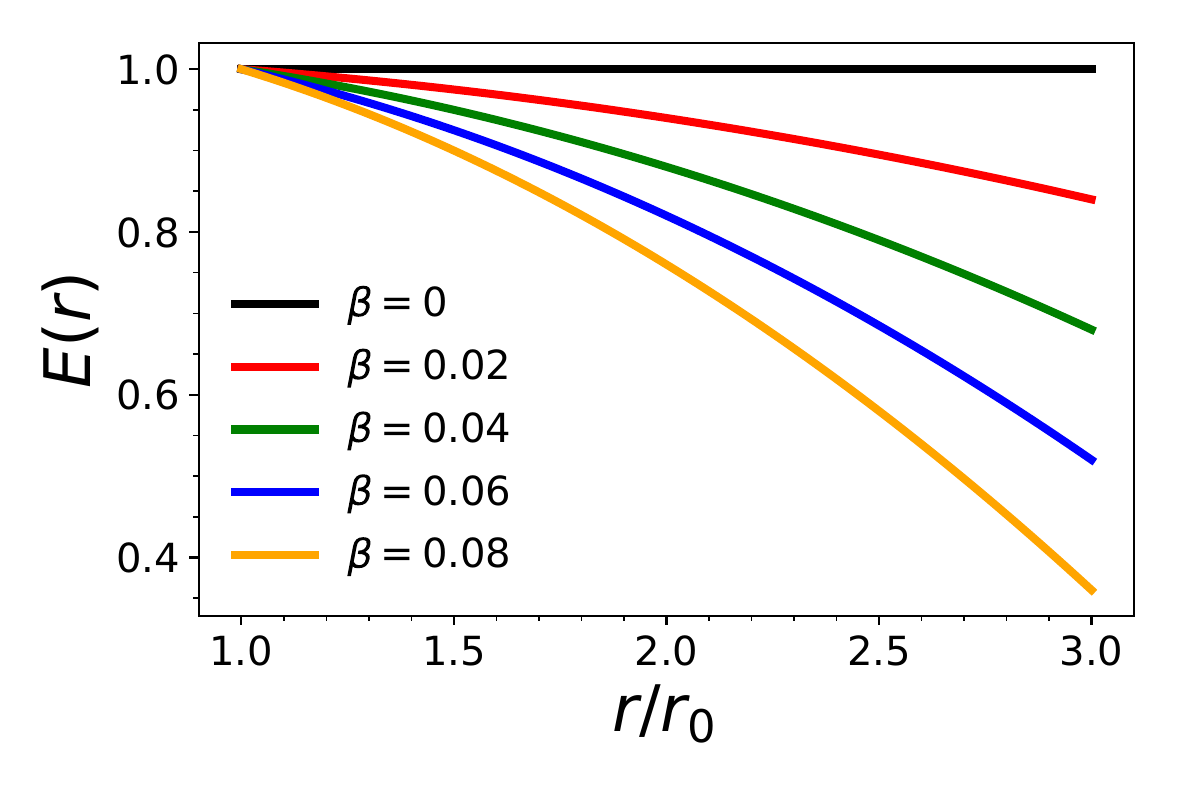}}
  \caption{Evaporation factor $E(r)$ plotted as a function of normalized radial distance $r/r_0$ with ${\beta}$ as a parameter. Five different values of ${\beta}{\sim}j{\pi}r_0^2/{\rho}Q_{in}$ are shown, representing different amounts of evaporation from the free liquid surface of the jet.}
\label{figure2}
\end{figure}

The radial velocity scale $V_*$ averaged over the film thickness from equation (3) equated with the variable rotational velocity scale (${\Omega}_0r$) results in a scaling relation for the liquid film thickness. This scaling physically signifies a rigid body type of motion for the entire liquid film.
  \begin{equation}
V_*{\sim}{\Omega}_0r{\sim}\frac{1}{2{\pi}rh(r)}[Q_{in} - {\frac{j{\pi}}{\rho}}(r^2-r_0^2)]
\end{equation}
The liquid film thickness scale becomes
\begin{equation}
{h(r){\sim}\frac{1}{2{\pi}{\Omega}_0r^2}[Q_{in} - {\frac{j{\pi}}{\rho}}(r^2-r_0^2)]}
\end{equation}
Substituting $Q_{in}={2{\pi}r_0h_0V_0}$ (refer to figure 1) we have
\begin{equation}
{h(r){\sim}\frac{Q_{in}}{2{\pi}{\Omega}_0r^2}[1 - {\frac{j{\pi}}{\rho{Q}_{in}}}(r^2-r_0^2)]}
\end{equation}

Let $E(r)$ denote the evaporation factor given by
    \begin{equation}
    E(r) = [1 - {\frac{j{\pi}r_0^2}{\rho{Q}_{in}}}({\frac{r^2}{r_0^2}}-1)]
  \end{equation}
  Introducing a nondimensional group to form a change of variables we have
  \begin{equation}
{\beta}={\frac{j{\pi}r_0^2}{\rho{Q}_{in}}}
\end{equation}
\begin{equation}
E(r) = [1 - {\beta}({\frac{r^2}{r_0^2}}-1)]
\end{equation}

\begin{figure}
  \centerline{\includegraphics[scale=0.9]{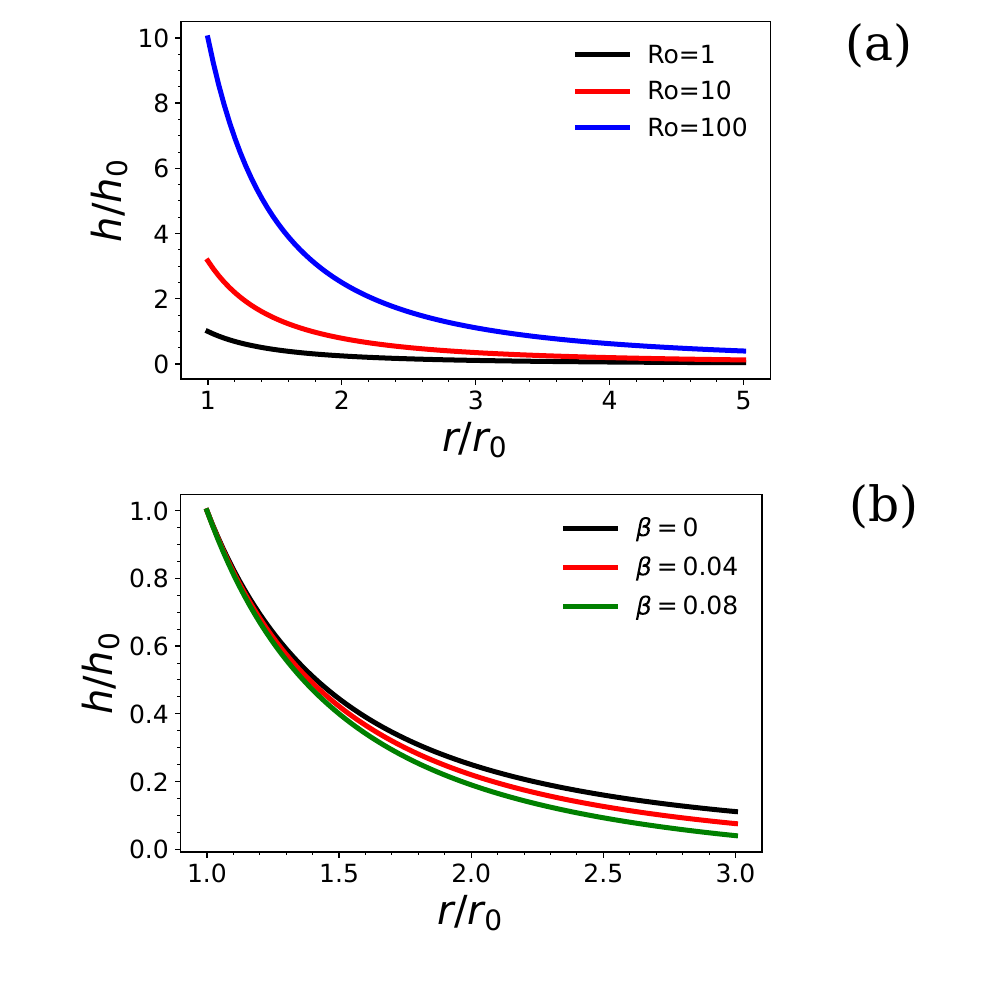}}
  \caption{Jet liquid film thickness profile plotted as a function of radial coordinate for inertia-rotation limit. (a) Dependence of the liquid profile with Rossby number as a parameter.
    (b) Dependence of the liquid profile with {${\beta}$} used as a parameter.}
\label{figure3}
\end{figure}

Equation (15) is plotted in figure 2.
The non-dimensional factor ${\beta}$ is the ratio of the evaporative mass loss through the liquid-vapor interface to the mass flow rate inside the control volume CV. ${\beta}=0$ represents zero evaporative flux through the liquid-air interface that is physically equivalent to a saturated environment (hundred percent relative humidity). The effect of ${\beta}$ and the radial coordinate on the evaporation factor $E(r)$ can be inferred from figure 2. $E(r)$ decreases as the radial coordinate and the non-dimensional factor ${\beta}$ increases. Using the above change of variables and representing the liquid film thickness as a fraction of a vertical length scale $h_0$ (mostly the vertical entrance scale in experiments reported in literature \cite{rice2005analysis} equation (14) can be rewritten as

\begin{equation}
\frac{h(r)}{h_0}{\sim}\left({\frac{r_0}{r}}\right)^2\sqrt{Ro}E(r)
\end{equation}

\begin{equation}
Ro=\frac{V_0^2}{{\Omega}_0^2r_0^2}
\end{equation}

Where $Ro$ is the Rossby Number. Equation (18) signifies the physical scaling of liquid film thickness under the effect of a rigid body rotation of the liquid film and inertia.
Figure 3(a) compares the liquid film thickness for different Rossby numbers and zero evaporative flux. The film thickness scaling was plotted using equation (18).
On increasing the Rossby number (i.e., decreasing rotational angular velocity ${\Omega}_0$), the liquid film thickness increases.
The effect of evaporation on the liquid film thickness can be understood from figure 3(b).
Figure 3(b) represents the liquid film thickness for a fixed Rossby number $Ro{\sim}1$ and a variable evaporative flux. The effect of evaporative flux becomes important as the radial coordinate increases.

\subsubsection{Inertial Inviscid Scaling with a constant scale for radial velocity (${\Omega}_0=0$)}
A different scaling limit exists for liquid film thickness by using a constant radial velocity scale ($V_*{\sim}V_0$) which physically resembles the situation of zero rotation of the plate.
Using equation (3) to assure integral mass conservation the radial velocity scales as
\begin{equation}
V_*{\sim}V_0{\sim}\frac{1}{2{\pi}rh(r)}[Q_{in} - {\frac{j{\pi}}{\rho}}(r^2-r_0^2)]
\end{equation}
Simplifying further and using $Q_{in}={2{\pi}r_0h_0V_0}$ we have 

\begin{equation}
\frac{h(r)}{h_0}{\sim}\frac{r_0}{r}[1 - {\frac{j{\pi}r_0^2}{\rho{Q}_{in}}}({\frac{r^2}{r_0^2}}-1)]
\end{equation}
Rewriting equation (21) using the evaporation factor $E(r)$ from equation (17) we have
\begin{equation}
    \frac{h(r)}{h_0}{\sim}\frac{r_0}{r}E(r)
\end{equation}
Comparing equation (18) and (22), we observe that the dependence on the evaporation factor is same (linear on $E(r)$). Increased evaporation flux corresponds to larger ${\beta}$ and smaller $E(r)$ and hence smaller liquid film thickness. However the liquid film thickness falls off faster ($1/r^2$) in case of rotation scaling compared to zero rotation (${1}/{r}$). 



\subsubsection{Viscous Rotation Scaling}
In practical applications involving high speed thin film liquid jets, viscous forces are important
due to high gradients of the velocity in a very thin region near the solid boundary.
In this section, the effects of rotation are now included to counteract the viscous effects.
Equating the viscous and rotation scaling from equation (9) we have
\begin{equation}
{\frac{{\nu}V_*}{z^2}}{\sim}{\Omega_0^2r}
\end{equation}
For evaluating the scale of liquid film thickness $h(r)$ we can scale $z{\sim}h(r)$
\begin{equation}
{\frac{{\nu}V_*}{{h}^2}}{\sim}{\Omega_0^2r}
\end{equation}

Using equation (3) in equation (24) we can establish a scale for the liquid film thickness incorporating the effects of viscosity, rotation and evaporation.
Therefore the scales of equation (24) can be rewritten as
  \begin{equation}
V_*{\sim}\frac{{\Omega}_0^2rh^2(r)}{\nu}{\sim}\frac{1}{2{\pi}rh(r)}[Q_{in} - {\frac{j}{\rho}}(r^2-r_0^2)]
\end{equation}

Simplifying and non-dimensionalizing $h(r)$ we have
\begin{equation}
  \frac{h(r)}{h_0}{\sim}\left(\frac{Ro}{Re}\right)^{1/3}\left(\frac{r_0}{h_0}\right)^{2/3}\left(\frac{r_0}{r}\right)^{2/3}[1 - {\frac{j{\pi}r_0^2}{\rho{Q}_{in}}}({\frac{r^2}{r_0^2}}-1)]^{1/3}
\end{equation}
where $Ro$ is the Rossby number
and
\begin{equation}
Re=\frac{V_0r_0}{\nu}
\end{equation}

Using the definition of $E(r)$ equation (26) can be rewritten as
\begin{equation}
  \frac{h(r)}{h_0}{\sim}\left(\frac{Ro}{Re}\right)^{1/3}\left(\frac{r_0}{h_0}\right)^{2/3}\left(\frac{r_0}{r}\right)^{2/3}[E(r)]^{1/3}
\end{equation}
Equation (28) shows the dependence of the normalized liquid film thickness on Rossby number, Reynolds number, and the evaporation factor, respectively ($Ro$, $Re$, $E(r)$). The liquid film thickness for viscous-rotation limit scales as $1/r^{2/3}$, which is different from the previous scalings derived before (refer to equation 18 and 22). Figures 4(a) and 4(b) show the dependence of the liquid film thickness profile on Reynolds number and Rossby number. Increasing the Rossby number (decreasing angular velocity ${\Omega}_0$) increases the liquid film thickness ($h(r)/h_0{\sim}Ro^{1/3}$) provided all the parameters are unchanged. The liquid film thickness is inversely proportional to Reynolds number ($h(r)/h_0{\sim}Re^{-1/3}$) as shown in figure 4(b). The dependence on the evaporation factor is non-linear ($h(r)/h_0{\sim}[E(r)]^{1/3}$). This scaling law is different from equation (18) and (22) where the dependence on the evaporation factor is linear. Note that the evaporation factor is always less than one (refer to figure 2, $E(r)<1$). Equation (28) has the same form as the closed-form solutions derived by Basu et al. for negligible inertia \cite{basu2006analysis}. Note that the correlations were derived from pure scaling logic, without any velocity profile assumption previously used in the literature. 

\begin{figure}
  \centerline{\includegraphics[scale=0.9]{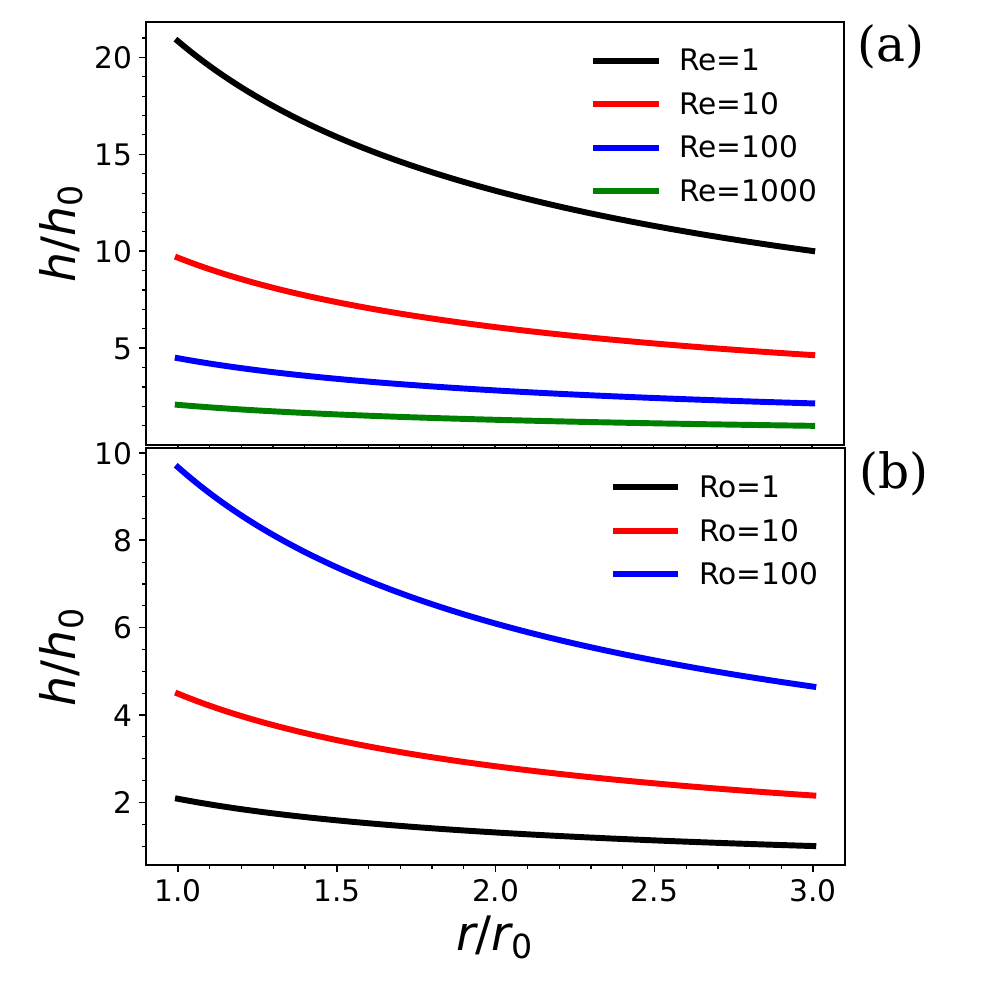}}
  \caption{Jet liquid film thickness profile plotted as a function of radial coordinate for viscous-rotation limit. (a) Dependence of the liquid profile with Reynolds number as a parameter.
    (b) Dependence of the liquid profile with Rossby number as a parameter.}
\label{figure4}
\end{figure}
   
Figure 5 compares the liquid film thickness scaling against some previous numerical and experimental data from literature \cite{rice2005analysis,thomas1990one}.
The graph shows the variation of liquid film thickness profile at a fixed Reynolds number with angular velocity (Rossby Number) as a parameter. Increasing the angular velocity decreases the liquid film thickness. The scalings predicted (refer to figure 4(b)) were in agreement with the experimental data reported in the literarture.
\begin{figure}
         \centerline{\includegraphics[scale=0.7]{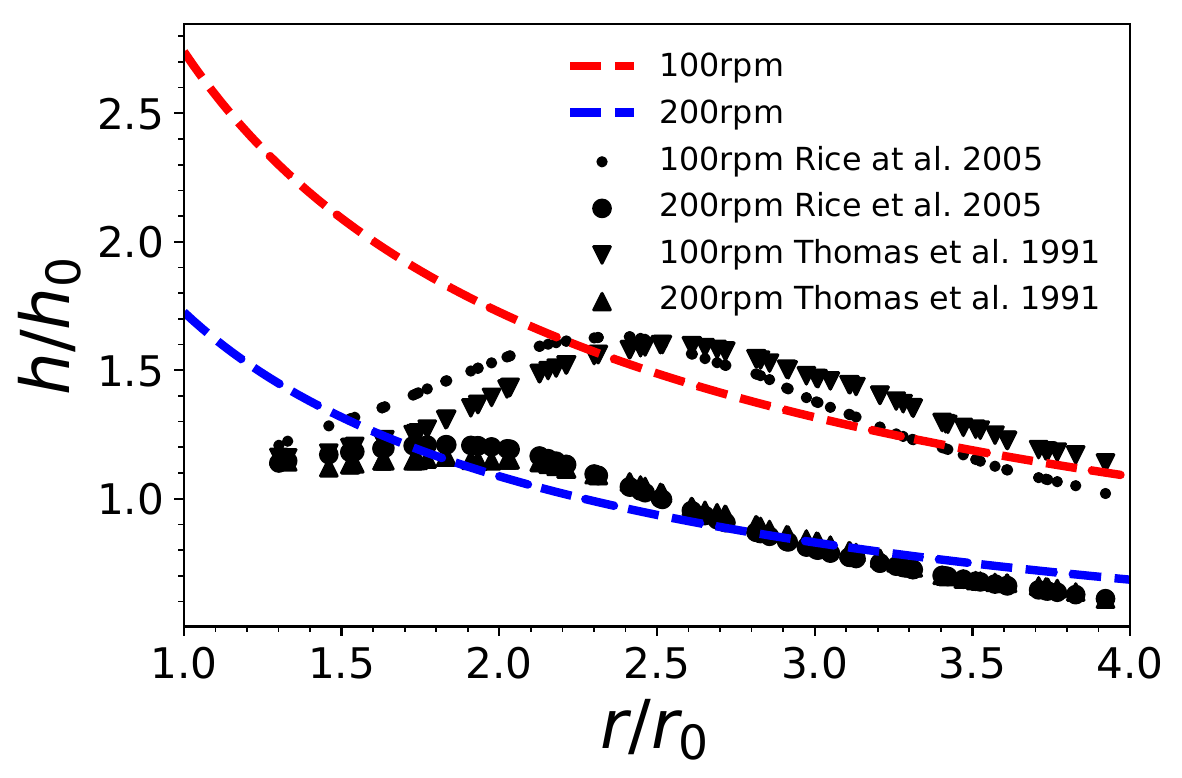}}
  \caption{Hydrodynamic liquid film thickness experimental and numerical comparison from literature using viscous-rotation scaling limit. The dashed line depicts the scales predicted using scaling theory.}
\label{figure5}
\end{figure}

\subsection{The average radial velocity field scaling}
In this section we derive the average radial velocity scales for the viscous rotation limit.
From equation (25) the average radial velocity scales as
\begin{equation}
V_*{\sim}\frac{{\Omega}_0^2rh^2(r)}{\nu}
\end{equation}
Defining $V_{\dagger}={\Omega}_0r_0$
and using equation (24) in equation (28) we have
\begin{equation}
\frac{V_*}{V_{\dagger}}{\sim}Ro^{2/3}Re^{1/3}\left(\frac{r_0}{h_0}\right)^{4/3}\left(\frac{r_0}{r}\right)^{1/3}E(r)^{2/3}
\end{equation}
Equation (30) deciphers the scales of the radial velocity on $Ro,Re,r$ and $E(r)$. The dependence on the radial coordinate is ${\sim}r^{-1/3}$. This shows the radial velocity decreases with radius. The physical condition is equivalent to the velocity field inside hydrodynamic boundary layers.
\subsection{Hydrodynamic Boundary Layer Thickness}
The heat transfer characteristics like the Nusselt number, the average temperature distribution of the liquid film can be understood in terms of boundary layer analysis. This section looks into the hydrodynamic boundary layer scalings, which will further lead to thermal boundary layer analysis in the later sections.
The radial flow velocity in the liquid rises very sharply from zero to the free surface jet velocity in a very thin boundary layer region.
Viscous forces balance inertial effects inside the boundary layer.
From the viscous scaling inside the boundary layer (using equation 9 and $z{\sim}{\delta}$), the boundary layer thickness scale for $r<r_v$ ($r_v$ defined below) is
\begin{equation}
  \frac{V_*^2}{r}{\sim}\frac{{\nu}V_*}{{\delta}^2}
\end{equation}
Using the scale of $(V_*)$ from equation (30) in equation (31), the boundary layer thickness scales as
\begin{equation} \frac{{\delta}}{r_0}{\sim}Re^{-2/3}Ro^{-1/12}\left(\frac{r_0}{h_0}\right)^{-2/3}\left(\frac{r}{r_0}\right)^{2/3}[E(r)]^{-1/3}
\end{equation}
\begin{figure*}
  \centerline{\includegraphics[scale=1.15]{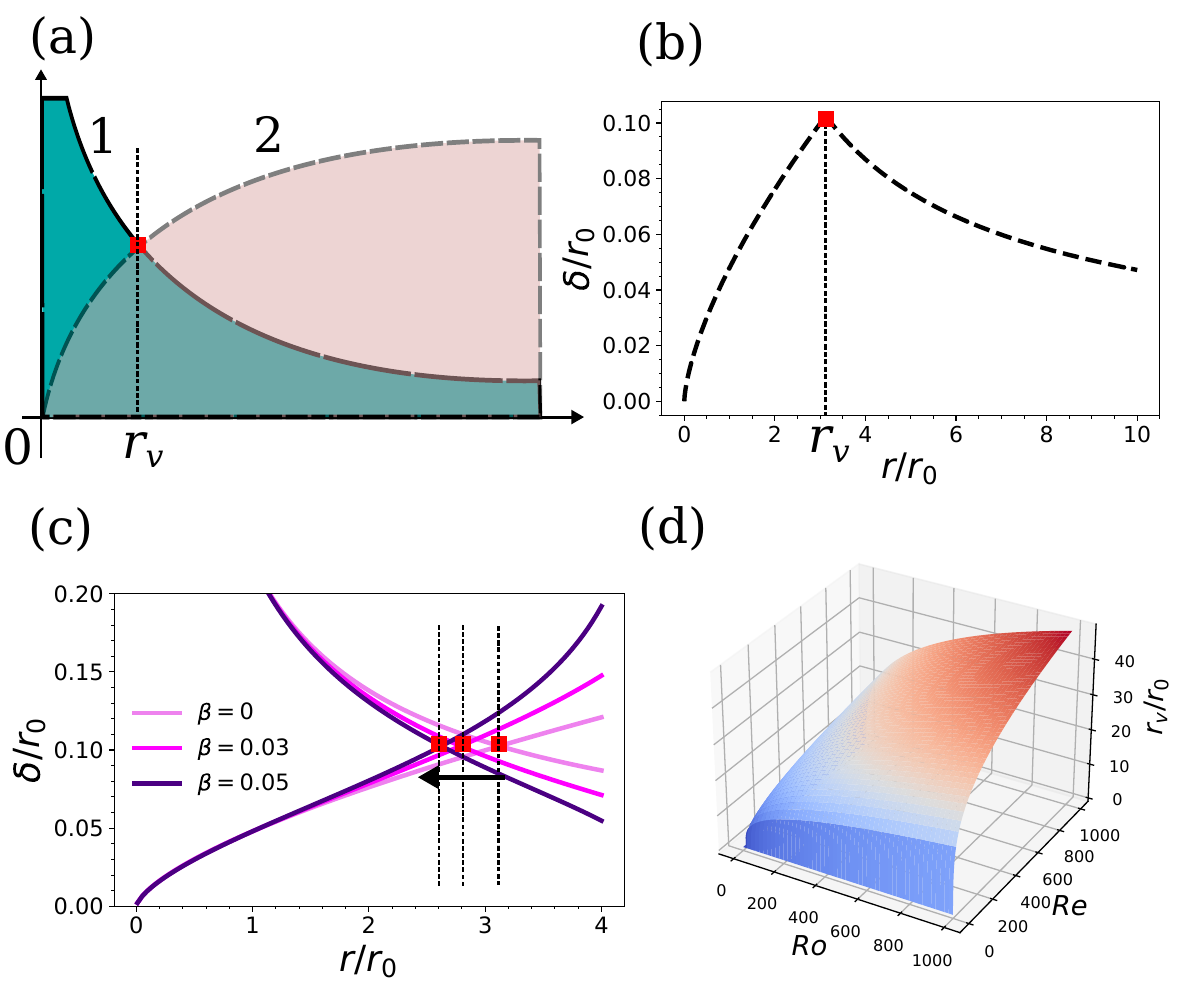}}
  \caption{(a) Schematic representation (not to scale) of the liquid film profile (shown by 1) and boundary layer profile (shown by 2). The intersection of profile 1 and 2 defines the radial location $r_v$ where the boundary layer profile reaches a maximum.
    (b) The combined piecewice boundary layer profile plotted for $Ro{\sim}Re{\sim}1$ and ${\beta}=0$.
    (c) The effect of ${\beta}$ on the combined boundary layer profile.
  (d) A surface plot showing the dependence of $r_v/r_0$ on Rossby number and Reynolds number for ${\beta}=0$.}
\label{figure6}
\end{figure*}

\begin{figure}  \centerline{\includegraphics[scale=0.8]{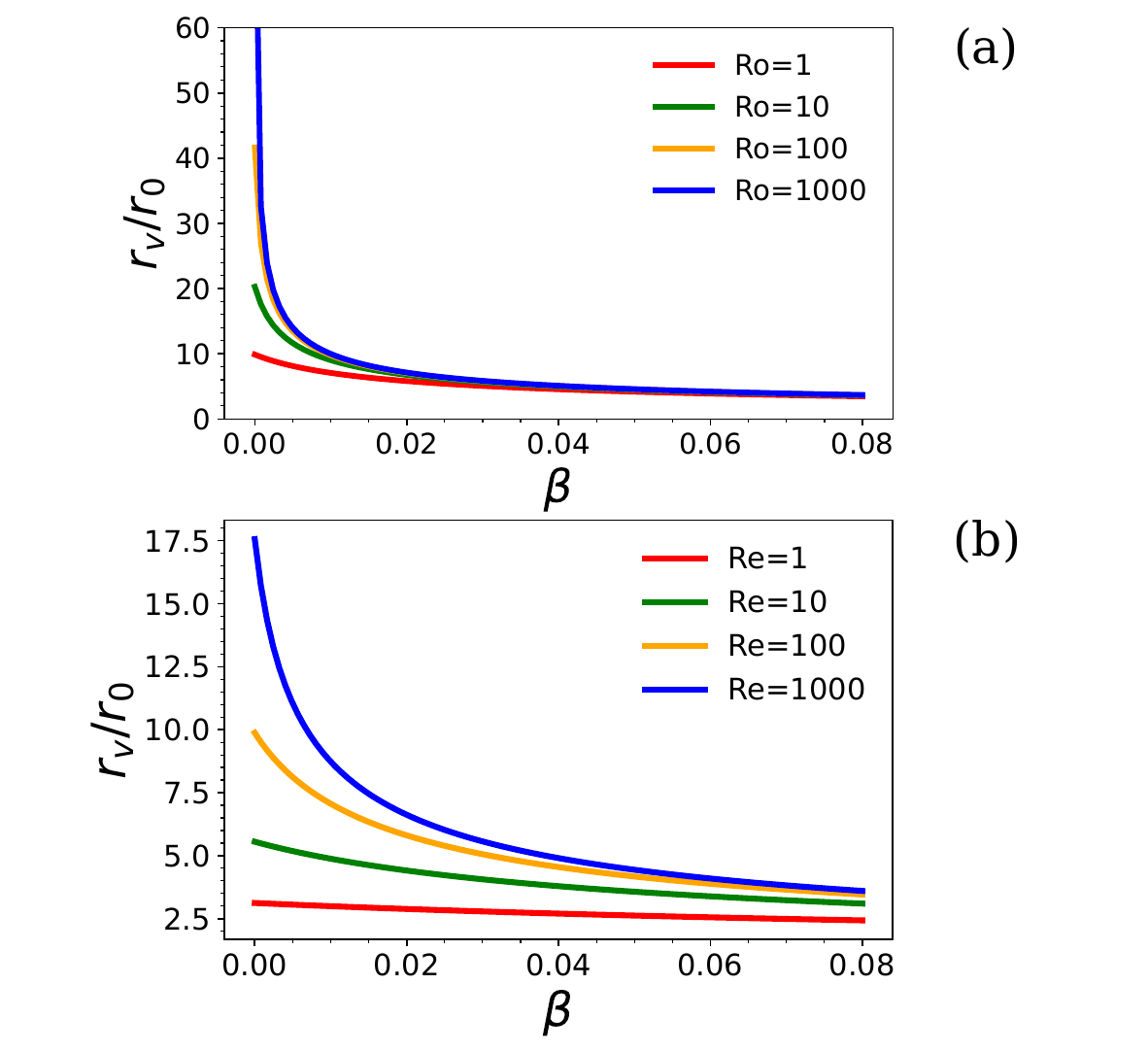}}
  \caption{(a) The variation of $r_v/r_0$ with respect to ${\beta}$ with Rossby number as the parameter for $Re{\sim}1$.
  (b) The variation of $r_v/r_0$ with respect to ${\beta}$ with Reynolds number as the parameter for $Ro{\sim}1$.}
\label{figure7}
\end{figure}  
The radial location at which the boundary layer profile peak is represented by ${r_v}$ (See figure 6(a)).
Equation (32) shows the dependence of the boundary layer thickness on the radial coordinate, Rossby number, Reynolds number, and evaporation factor for $r<r_v$
For $r>r_v$, the boundary layer thickness scales as the liquid film thickness for the viscous rotation scaling given by

\begin{equation}
\frac{\delta}{r_0}{\sim}\frac{h(r)}{r_0}{\sim}\left(\frac{Ro}{Re}\right)^{1/3}\left(\frac{h_0}{r_0}\right)^{1/3}\left(\frac{r_0}{r}\right)^{2/3}E(r)^{1/3}
\end{equation}

Equation (32) and (33) defines the boundary layer thickness scales in a piecewise manner. This piecewise function ${\delta}$ is plotted as a function of radial coordinate for $Ro{\sim}Re{\sim}1$ and ${\beta}=0$ in figure 6(b). Comparing equation (32) with (33) depicts that the effects of Rossby number and evaporation factor on liquid film thickness is reversed before and after the critical radius $r_v$. 
   \subsection{Scaling of the radial location $r_v$}
   The radial location $r_v$ is the distance where the viscous boundary layer from the plate meets the outer liquid film thickness profile in the viscous rotation scaling (refer to figure 6(a) and 6(b)). The existence of $r_v$ is only valid for cases where the thermal boundary is much much smaller than 1, i.e., ${\delta}/{{\delta}_T}>>1$. Equating the scales of equation (32) and (33) and solving for $r_v$ in case of $E(r)=1$ (i.e. No evaporation) we have
       \begin{equation}
           \frac{r_v}{r_0}{\sim}\left(\frac{r_0}{h_0}\right)^{1/4}Ro^{5/16}Re^{1/4}
         \end{equation}

         The effect of evaporation on the radial coordinate $r_v$ can be inferred from figure 6(c). Increased evaporation flux rate reduces $r_v$. It can be observed by the shift in the intersection position (shown as a red solid square in figure 6(c)) of the initial increasing boundary layer profile and the decreasing liquid film thickness profile. 
         Equation (35) shows the dependence of $r_v$ on $Ro$ and $Re$ for ${\beta}=0$. The dependence is visualized as a surface map shown in figure 6(d). 
       
For $0<E(r)<1$, the scale of $r_v/r_0$ can be solved explicitly by keeping the terms of $E(r)$ after equating the scales of equation (32) and (33).
\begin{equation}
\begin{split}
Re^{-2/3}Ro^{-1/12}\left(\frac{r_0}{h_0}\right)^{-2/3}\left(\frac{r_v}{r_0}\right)^{2/3}[E(r_v)]^{-1/3}{\sim}\\Ro^{1/3}Re^{-1/3}\left(\frac{r_0}{h_0}\right)^{-1/3}\left(\frac{r_v}{r_0}\right)^{-2/3}[E(r_v)]^{1/3}
\end{split}
\end{equation}
Equation (35) can be simplified and solved explicitly for ${r_v/r_0}$ in this case (Note algebraically this is not always possible).
\begin{equation}
\frac{r_v}{r_0}{\sim}\sqrt{\frac{1+{\beta}}{{\beta}+F(Ro,Re,r_0,h_0)}}
\end{equation}
where 
\begin{equation}
F(Ro,Re,r_0,h_0)=Ro^{-5/8}Re^{-1/2}\left(\frac{r_0}{h_0}\right)^{-1/2}
\end{equation}
Equation (36) is plotted in figure 7(a) and 7(b). Figure 7(a) depicts the dependence of the critical radius $r_v$ on ${\beta}$ with Rossby number as the parameter. Similarly, figure 7(b) depicts the dependence of the critical radius $r_v$ on ${\beta}$ with Reynolds number as the parameter. Both figure 7(a) and 7(b) shows that the critical length scale $r_v$ decreases with incresing ${\beta}$ i.e., increased evaporation rate.

\subsection{Thermal boundary layer thickness and Nusselt Number for ${\delta}/{{\delta}_T}{\sim}1$}
In this and the following two sections we discuss the thermal characteristics for various limiting cases. We calculate the Nusselt number scaling for ${\delta}/{\delta}_T{\sim}1$, ${\delta}/{\delta}_T>1$ and ${\delta}/{\delta}_T<1$.
As discussed in the previous section, the hydrodynamic boundary layer thickness scalings can be defined in a piecewise manner using two limiting conditions. Following a similar line of reasoning for thermal boundary layer and Nusselt number, we have two regimes $r<r_v$ and $r>r_v$.
For $r<r_v$, the thermal boundary layer height scales as the hydrodynamic boundary layer (${\delta}{\sim}{\delta}_T$). The thermal boundary layer thickness from equation (32) can be written, therefore as
\begin{equation}
\frac{\delta_T}{r_0}{\sim}\frac{{\delta}}{r_0}{\sim}Re^{-2/3}Ro^{-1/12}\left(\frac{r_0}{h_0}\right)^{-2/3}\left(\frac{r}{r_0}\right)^{2/3}[E(r)]^{-1/3}
\end{equation}
Simplifying further the thermal boundary layer thickness written in terms of non-dimensional parameters like Rossby number ($Ro$) and Reynolds number ($Re$) is
\begin{equation}
  \frac{{\delta}_T}{r_0}{\sim}Re^{-1/2}Ro^{1/4}\left(\frac{r}{r_0}\right)^{1/2}\left(\frac{V_*}{V_{\dagger}}\right)^{-1/2}
\end{equation}
Using equation (30) in equation (39) for $V_*/V_{\dagger}$
\begin{equation}
\frac{{\delta}_T}{r_0}{\sim}Re^{-2/3}Ro^{-1/12}\left(\frac{r_0}{h_0}\right)^{-2/3}\left(\frac{r}{r_0}\right)^{2/3}[E(r)]^{-1/3}
\end{equation}

The Nusselt number scale is the reciprocal of the non dimensional thermal boundary layer thickness scale
\begin{equation}
Nu{\sim}\frac{r_0}{{\delta}_T}{\sim}Re^{2/3}Ro^{1/12}\left(\frac{r_0}{h_0}\right)^{2/3}\left(\frac{r}{r_0}\right)^{-2/3}[E(r)]^{1/3}
\end{equation}
For $r>r_v$ the boundary layer length scales as the liquid film thickness (refer to equation 33)
\begin{equation}
\frac{\delta}{r_0}=\frac{{\delta}_T}{r_0}{\sim}\left(\frac{h_0}{r_0}\right)^{1/3}\left(\frac{Ro}{Re}\right)^{1/3}\left(\frac{r_0}{r}\right)^{2/3}[E(r)]^{1/3}
\end{equation}

The Nusselt number scaling can be rewritten as
\begin{equation}    Nu{\sim}\frac{r_0}{{\delta}_T}{\sim}\left(\frac{h_0}{r_0}\right)^{-1/3}\left(\frac{Ro}{Re}\right)^{-1/3}\left(\frac{r_0}{r}\right)^{-2/3}[E(r)]^{-1/3}
\end{equation}
Figure 8(a) shows the variation of Nusselt number limiting scales with radial coordinate for $Ro{\sim}1$, $Re{\sim}1$, $Pr{\sim}7$ and ${\beta}$ as the parameter. For $0<r/r_0<r_v/r_0$, Nusselt number decreases with radial coordinate due to an increasing thermal boundary layer thickness scale. The Nusselt number scales increases with radial coordinate $r/r_0>r_v/r_0$ due to subsequent reduction of liquid film thickness. Equation (41) and (43) together shows the dependence on Rossby number, Reynolds number and the evaporation factor in a piecewice manner for $r<r_v$ and $r>r_v$ respectively. The dependence of Nusselt number $Nu$ on evaporation factor $E(r)$ is reversed before and after the critical length scale $r_v$. It can be inferred from figure 8(a) that the increased evaporation flux (higher value of ${\beta}$) causes an increase in Nusselt number. 

\begin{figure*}  \centerline{\includegraphics[scale=1]{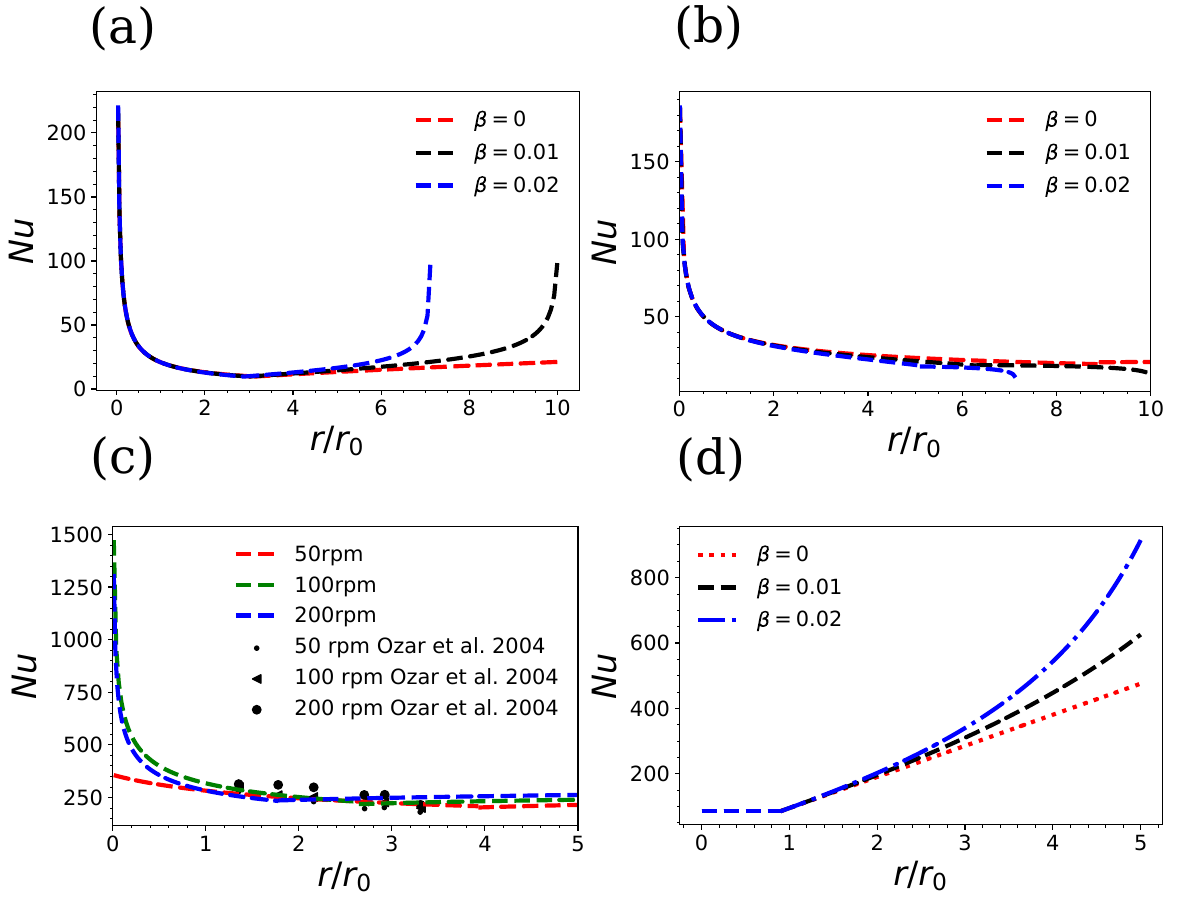}}
  \caption{(a) Nusselt number scales plotted as a function of radial coordinate for ${\delta}{\sim}{\delta}_T$ for $Ro{\sim}Re{\sim}1$, $Pr=7$, and ${\beta}$ as the parameter.
    (b) Nusselt number scales plotted as a function of radial coordinate for ${\delta}{>>}{\delta}_T$ for $Ro{\sim}Re{\sim}1$, 
    $Pr{\sim}7$, and ${\beta}$ as the parameter.
    (c) Nusselt number scales comparison with experimental and numerical data available from literature. The dashed lines represents the scales predicted for Nusselt number.
    (d) Nusselt number scales plotted as a function of radial coordinate for ${\delta}{<<}{\delta}_T$ for $Re{\sim}10000$,  $Pr=0.1$, and ${\beta}$ as the parameter.
  }
\label{figure8}
\end{figure*}

  \subsection{Thermal boundary layer thickness and Nusselt Number for ${\delta}/{{\delta}_T}>>1$}
  This section develops the thermal characteristics scales for conditions where the hydrodynamic boundary layer thickness is very much greater than the thermal boundary layer thickness. From equation (22), we have the scaling form of the thermal energy equation inside the thin thermal boundary layer region.
  
\begin{equation}
  \frac{V_r{\Delta}T}{r_0},\frac{V_z{\Delta}T}{\delta_T}{\sim}\frac{{\alpha}{\Delta}T}{{\delta}_T^2}
\end{equation}
  Since the thermal boundary layer is wholly immersed inside the hydrodynamic boundary layer, there exists a gradient of radial velocity in the z-direction inside the thermal boundary layer. The radial velocity scale inside the thermal boundary layer is therefore given by
  \begin{equation}
  V_r{\sim}\frac{{V_*}{\delta}_T}{\delta}
  \end{equation}
  From the scaling form of equation (4), the relationship between the radial velocity $V_r$ and the axial velocity $V_z$ is
  \begin{equation}
\frac{V_r}{r_0}{\sim}\frac{V_z}{{\delta}_T}
  \end{equation}
  Using equation (45) in equation (46), the axial velocity scales as
   \begin{equation}
   V_z{\sim}\frac{{\delta}_T{V}_r}{r_0}{\sim}\frac{V_*{\delta}_T^2}{r_0{\delta}}
 \end{equation}
 Using the scales of $V_r$ and $V_z$ from equation (45) and (47) respectively in equation (44) the scaling form of the thermal energy equation becomes
 \begin{equation}
\frac{V_*{\delta}_T}{r_0{\delta}}, \frac{V_*{\delta}_T^2}{r_0{\delta}{\delta}_T}{\sim}\frac{{\alpha}}{{\delta}_T^2}
 \end{equation}
 The first and second term scales in equation (48) are equivalent and therefore equation (48) can be rewritten as
   \begin{equation}
   \frac{V_*}{r_0}\frac{{\delta}_T}{\delta}{\sim}\frac{\alpha}{{\delta}_T^2}
   \end{equation}
   
   \begin{equation}
   \left(\frac{{\delta}_T}{r_0}\right)^3{\sim}\frac{\alpha}{V_*r_0}\frac{\delta}{r_0}
 \end{equation}
 Equation (50) can be simplified using non-dimensional numbers and rewritten as
 \begin{equation}
\left(\frac{{\delta}_T}{r_0}\right)^3{\sim}Pe^{-1}Ro^{1/2}\left(\frac{V_{\dagger}}{V_*}\right)\left(\frac{\delta}{r_0}\right)
 \end{equation}
 where $Pe=RePr=V_0r_0/{\alpha}$ is the Peclet number.
   The boundary layer thickness for $r<r_T$ ($r_T$ is the radial coordinate at which the thermal boundary layer meets the liquid film thickness) is given by equation (32). Substituting the value of ${\delta}/r_0$ in equation (51) and solving for ${{\delta}_T}/{r_0}$
   \begin{equation}
\frac{{\delta}_T}{r_0}{\sim}Pr^{-1/3}Re^{-2/3}Ro^{-1/12}\left(\frac{r_0}{h_0}\right)^{-2/3}\left(\frac{r}{r_0}\right)^{1/3}[E(r)]^{-1/3}
\end{equation}

  Hence the scale for the Nusselt number can be calculated using the above equation
  \begin{equation}
      Nu{\sim}\frac{r_0}{{\delta}_T}{\sim}Pr^{1/3}Re^{2/3}Ro^{1/12}\left(\frac{r_0}{h_0}\right)^{2/3}\left(\frac{r}{r_0}\right)^{-1/3}[E(r)]^{1/3}
  \end{equation}
  For $r>r_T$, substituting the scale of the hydrodynamic boundary layer thickness from equation (33) into equation (51) the thermal boundary layer scales as
 
\begin{equation}
 \left(\frac{{\delta}_T}{r_0}\right){\sim}Pr^{-1/3}Ro^{1/18}Re^{-5/9}\left(\frac{r_0}{h_0}\right)^{23/45}\left(\frac{r_0}{r}\right)^{1/9}[E(r)]^{-1/9}   
\end{equation}

Hence the Nusselt number scales as
\begin{equation}
    Nu{\sim}Pr^{1/3}Ro^{-1/18}Re^{5/9}\left(\frac{r_0}{h_0}\right)^{-23/45}\left(\frac{r_0}{r}\right)^{-1/9}[E(r)]^{1/9}
\end{equation}
Equation (53) alongwith (55) defines Nusselt number scales in a piecewice manner. Figure 8(b) depicts the scales of Nusselt number in a graphical way for water (having $Pr{\sim}7$, $Ro{\sim}Re{\sim}1$ and ${\beta}$ as the parameter). The effect of evaporation (${\beta}$) on Nusselt number in this case is negligible as can be inferred from figure 8(b) which different from the case when ${\delta}_T{\sim}{\delta}$ (refer to figure 8(a)). The Nusselt number decreases with radial coordinate continuously. This can be attributed to the fact that the thermal boundary layer is very small compared to the liquid film thickness and just keeps growing reducing the Nusselt number.
Figure 8(c) shows the Nusselt number comparison with some previous references avialiable from literature \cite{ozar2004experiments}. The scales predicted from analysis is consistent with the experimental data.

\subsection{Scaling of $r_T$ for ${\delta}/{{\delta}_T}>>1$}
    The radial coordinate $r_T$ is the distance from the axis of rotation where the thermal boundary layer intersects the liquid film thickness profile. The scale for $r_T$ can be deduced by equating the scales from equation (52) and equation (33). However, unlike the case for ${\delta}_T{\sim}{\delta}$ the scale for $r_T$ cannot evaluated algebraically. 
    The equation relating the scales of $r_T$ with other parameters is
    \begin{equation}
    \left(\frac{r_T}{r_0}\right)^{-3/2}[E(r_T)]{\sim}G(Ro,Re,Pr,r_0,h_0)
    \end{equation}
    where 
    \begin{equation}
    G(Ro,Re,Pr,r_0,h_0)=Ro^{-5/8}Re^{-1/2}Pr^{-1/2}\left(\frac{r_0}{h_0}\right)^{-1/2}
    \end{equation}
    We have solved equation (56) for $r_T/r_0$ graphically using three different values of ${\beta}=0,0.01,0.02$.
Figure 9(a) shows the graphical solution for $r_T$. It can be inferred from figure 9(a) that the critical length scale $r_T$ decreases with increase in evaporation flux represented by ${\beta}$. The vertical axis $H(r_T/r_0)$ in figure 9(a) is a function of ${r_T/r_0}$ with ${\beta}$ as a parameter.
\begin{equation}
H(r_T/r_0)=\frac{1-{\beta}}{(r_T/r_0)^{3/2}}-{\beta}(r_T/r_0)^{1/2}
\end{equation}
From figure 9(a) the values of ${r_T}/r_0$ can be evaluated. 
${r_T}/r_0$ calculated are $8.9$ for ${\beta}=0$, $6.3$ for ${\beta}=0.01$ and $5.1$ for ${\beta}=0.02$.
\begin{figure}  \centerline{\includegraphics[scale=0.8]{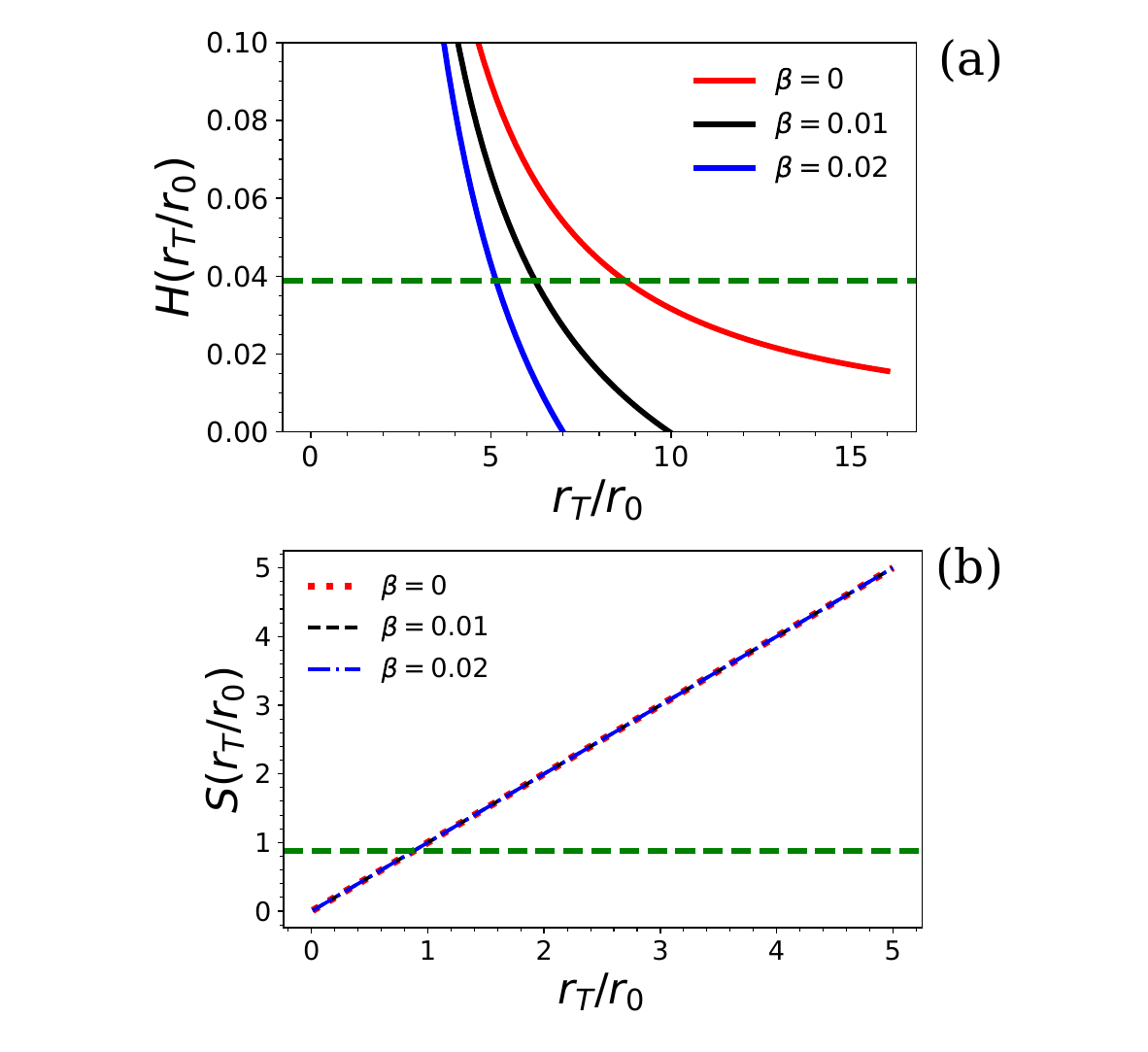}}
  \caption{(a) Graphical solution of $r_T/r_0$ for ${\delta}/{{\delta}_T}>>1$ with ${\beta}$ as the parameter.
(b) Graphical solution of $r_T/r_0$ for ${\delta}/{{\delta}_T}<<1$ with ${\beta}$ as the parameter for $Re{\sim}1000$. 
  }
\label{figure9}
\end{figure}  

\subsection{Thermal boundary layer thickness and Nusselt Number for ${\delta}/{{\delta}_T}<<1$}
This section deals with the regime where the thermal boundary layer is much larger than hydrodynamic boundary layer thickness. Practically liquid metals fall in this category.
The radial velocity $V_r$ scales as $V_*$. From the scaling of the differential form of continuity equation (5) the velocity scale in the axial direction is

\begin{equation}
    V_z{\sim}\frac{{\delta}_T}{r_0}V_*
\end{equation}
Using the scales of $V_r$ and $V_z$ in equation (11) the scaling equivalent of the thermal energy equation becomes
\begin{equation}
\frac{V_*}{r_0},\frac{V_*{\delta}_T}{r_0{\delta}_T}{\sim}\frac{\alpha}{{\delta}_T^2}
\end{equation}
Simplifying equation (60) and solving for the thermal boundary layer thickness we have
\begin{equation}
\frac{{\delta}_T}{r_0}{\sim}Pe^{-1/2}Ro^{1/4}\left(\frac{V_{\dagger}}{V_*}\right)^{1/2}
\end{equation}

Using the scale of $V_*{\sim}V_0$ and using $V_{\dagger}/V_0{\sim}Ro^{-1/2}$, equation (61) can be rewritten for $r<r_T$
\begin{equation}
  \frac{{\delta}_T}{r_0}{\sim}Pe^{-1/2}{\sim}Pr^{-1/2}Re^{-1/2}
\end{equation}
The Nusselt number scale for $r<r_T$ is therefore the reciporcal of equation (62)
\begin{equation}
  {Nu}{\sim}Pr^{1/2}Re^{1/2}
\end{equation}
For $r>r_T$, the thermal boundary layer scales as the liquid film thickness as shown before in equation (22). The radial coordinate $r_T$ is the distance from the central axis, where the thermal boundary layer meets the liquid film thickness profile. For $r>r_T$ the Nusselt number scales as
\begin{equation}
  Nu{\sim}\left(\frac{r}{r_0}\right)\left(\frac{r_0}{h_0}\right)[E(r)]^{-1}
\end{equation}
Figure 8(d) depicts the Nusselt number as decribed by equations (63) and (64) for $Re{\sim}10000$, $Pr=0.1$ and ${\beta}$ as the parameter. The Nusselt number remains constant upto a critical radius $r_T$, beyond which the Nusselt number increases with radial coordinate. The evaporation is important at larger radial coordinate $r/r_0>2$. It can be further inferred from figure 8(d) that increased evaporation flux (means higher ${\beta}$) increases the Nusselt number. Note due to a very thin hydrodynamic boundary layer, Nusselt number does not dependent on Rossby number.

\subsection{Scaling of $r_T$ for ${\delta}/{{\delta}_T}<<1$}
The scaling for $r_T$ in this regime can be developed by equating the scales of thermal boundary layer height profile with liquid film thickness profile. Equating scales of equation (62) with equation (33) we have
       \begin{equation}
 \frac{r_T}{r_0}\frac{1}{E(r_T)}{\sim}
 Pr^{1/2}Re^{1/2}\left(\frac{h_0}{r_0}\right)
         \end{equation}
 As before, we have solved equation (65) graphically for $r_T/r_0$. The solution is represented in figure 9(b). $S(r_T/r_0)$ represents a function of $r_T/r_0$ and ${\beta}$.
 \begin{equation}
 S(r_T/r_0)=\frac{r_T/r_0}{1-{\beta}((r_T/r_0)^2-1)}
 \end{equation}
 Figure 9(b) represents an interesting feature that critical radius $r_T$ in this regime is independent of evaporation flux.
 The critical value of $r_T/r_0$ is $0.9$ for $Re{\sim}1000$ and $Pr{\sim}0.1$.
         The scaling laws for the thermal boundary layer are used to derive the average liquid jet temperature scale. The two different kinds of boundary conditions, isothermal plate and constant heat flux case, are discussed in the following sections.
         \subsection{Thin film average temperature for isothermal plate}
         The liquid thin film average temperature can be evaluated by using the scales of thermal boundary layer thickness.
Applying integral conservation of thermal energy across the CV (refer to figure 1)
  \begin{equation}
  \begin{split}
{\rho}Q_{in}c_pT_i + \frac{k(T_0-T(r))}{{\delta}_T(r)}{\sim}{\rho}V_r(r)2{\pi}rh(r)c_pT(r)\\+ j{\pi}(r^2-r_0^2)h_{fg}
\end{split}
\end{equation}
where $c_p$ is the specific heat capacity of the liquid, ${\rho}$ is the density of the liquid, $T_i$ is the temperature of the jet at inlet entering the control volume, $k$ is the thermal conductivity of the liquid, $T_0$ is the temperature of the rotating isothermal plate, $T(r)$ is the temperature of the liquid film averaged over the thermal boundary layer thickness, $h_{fg}$ is the latent heat of vaporization. 
From integral continuity equation (3) we have
\begin{equation}
rV_r(r)h(r){=}\frac{1}{2{\pi}}[Q_{in}-\frac{j{\pi(r^2-r_0^2)}}{\rho}]
\end{equation}
Substituting equation (68) in equation (67) the thermal energy balance equation scales as   
\begin{equation}
\begin{split}
{\rho}Q_{in}c_pT_i+\frac{k(T_0-T(r))}{{\delta}_T(r)}{\sim}{\rho}c_p[Q_{in}-\frac{j{\pi}(r^2-r_0^2)}{\rho}]T(r)\\+j{\pi}(r^2-r_0^2)h_{fg}
\end{split}
\end{equation}
where
  $A={\rho}Q_{in}c_pT_i$, $f(r)={\rho}c_p[Q_{in}-\frac{j{\pi}(r^2-r_0^2)}{\rho}]$,\\ $g(r)=j{\pi}(r^2-r_0^2)h_{fg}$.
  \begin{equation}
    A+\frac{k(T_0-T(r))}{{\delta}_T(r)}{\sim}f(r)T(r)+g(r)
\end{equation}
The first term on the left hand $A$ signifies the internal energy entering the control volume. The second term denotes the average heat flux entering the control volume from the heated rotating plate
while $f(r)$ and $g(r)$ are related to the effects of evaporative flux on the average liquid film temperature.

\begin{equation}
  T(r)[f(r)+\frac{k}{{\delta}_T(r)}]{\sim}A+\frac{kT_0}{{\delta}_T(r)}-g(r)
\end{equation}

\begin{equation}
T(r){\sim}\frac{A+\frac{kT_0}{{\delta}_T(r)}-g(r)}{f(r)+\frac{k}{{\delta}_T(r)}}
\end{equation}
Most importantly, the average liquid film temperature scale depends on the temperature of the liquid jet entering the control volume and the thermal boundary layer thickness scale. Figure $10(a)$ shows the average temperature variation with radial coordinate for different ${\beta}$ as described by equation (72). Increasing beta (evaporation flux) increases the average temperature field. The maximum temperature peak also shifts towards the left. Figure 9(a) is plotted for water with inlet temperature of $T_{i}=298K$ and constant plate temperature of $T_0{=}313K$. 
\begin{figure}  \centerline{\includegraphics[scale=0.8]{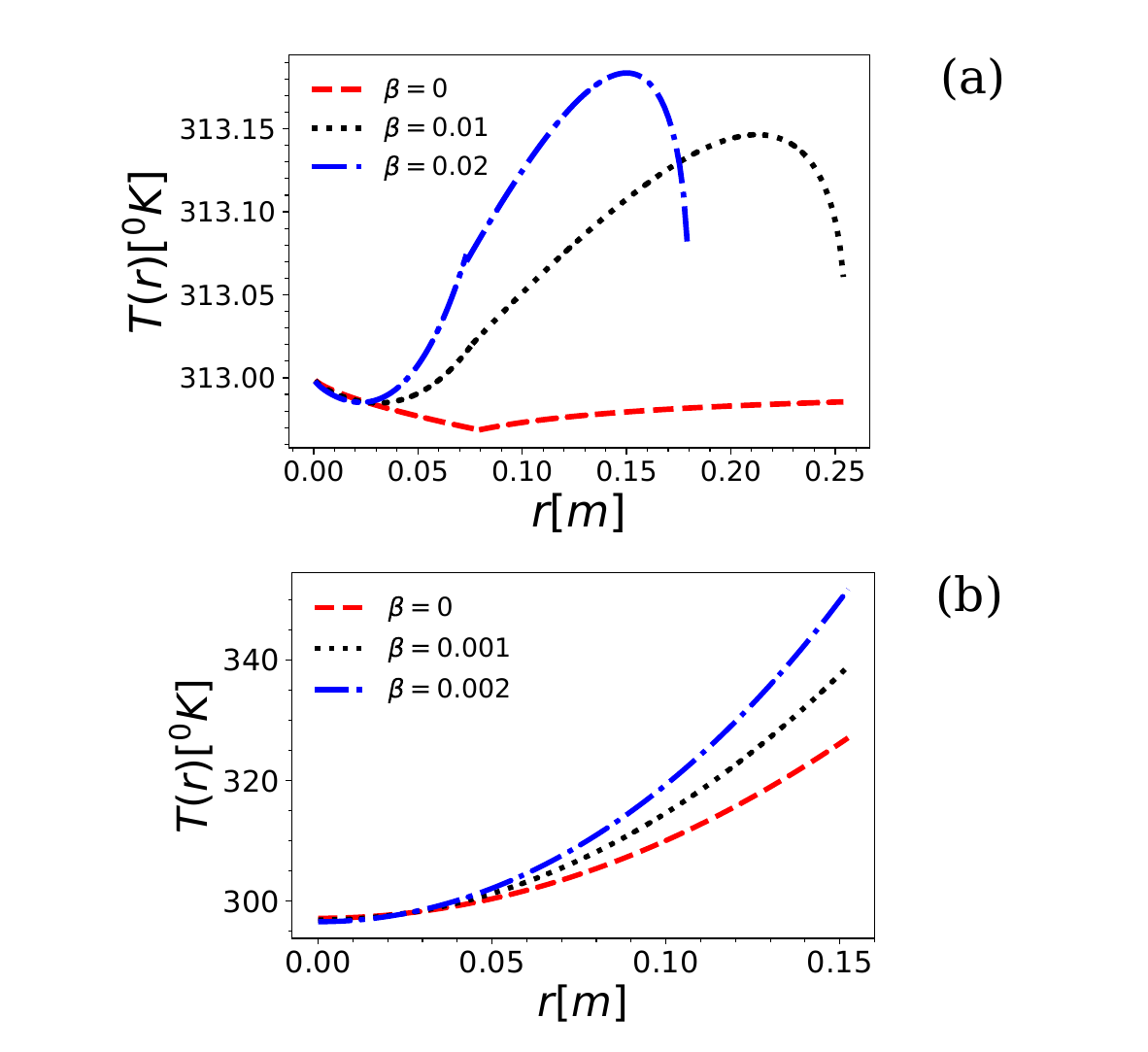}}
  \caption{(a) Average liquid film temperature as a function of radial coordinate with ${\beta}$ as the parameter for isothermal plate. The plate temperature is at $T_0{=}313K$ and the liquid inlet temperature entering the control volume is $T_i{=}298K$.
(b) Average liquid film temperature as a function of radial coordinate with ${\beta}$ as the parameter for constant heat flux case. The inlet temperature of the jet is $T_i{=}298K$ and the constant heat flux at the plate is $q^{''}=2{\times}10^5W/m^2$.
  }
\label{figure10}
\end{figure}  

\subsection{Thin film average temperature for constant heat flux case}
Following a similar line of reasoning as for constant temperature conditions, the average liquid film temperature scale for constant heat flux is derived in this section.
Applying integral conservation of thermal energy (refer to figure 1)
  \begin{equation}
  \begin{split}
{\rho}Q_{in}c_pT_i + q^{''}{\pi}(r^2-r_0^2){\sim}{\rho}V_r(r)2{\pi}rh(r)c_pT(r)\\+ j{\pi}(r^2-r_0^2)h_{fg}
\end{split}
\end{equation}
where $q^{''}$ is the constant heat flux at the surface of the rotating heated plate. 
Using integral continuity equation (3)

\begin{equation}
\begin{split}
{\rho}Q_{in}c_pT_i+q^{''}{\pi}(r^2-r_0^2){\sim}{\rho}c_p[Q_{in}-\frac{j{\pi}(r^2-r_0^2)}{\rho}]T(r)\\+j{\pi}(r^2-r_0^2)h_{fg}
\end{split}
\end{equation}
The average liquid film temperature scales as
\begin{equation}
T(r){\sim}\frac{{\rho}Q_{in}c_pT_i+q^{''}{\pi}(r^2-r_0^2)-j{\pi}(r^2-r_0^2)h_{fg}}{{\rho}c_p(Q_{in}-\frac{j{\pi}(r^2-r_0^2)}{\rho})}
\end{equation}
Comparing equation (72) with equation (75) shows the dependence of the average temperature field on the thermal boundary layer thickness scale for constant temperature boundary conditions.
Further the average radial temperature field is directly proportional to the inlet temperature $T_i$ and the heat flux $q^{''}$ as can be inferred from equation (75). The average liquid film temperature scale is plotted using equation (75) in figure 10(b). The effect of evaporation is understood by visualizing the dependence of the average temperature on ${\beta}$. The average temperature of the liquid film increases with radial coordinate, and increases with increasing evaporation flux.
This is in accordance with the constant heat flux boundary conditions, where the average temperature increases monotonically with radial coordinate.
\section{Conclusion}
We provided an integrodifferential scaling analysis to analyze the steady laminar free jet impacting a rotating hot plate. 
Integral mass conservation scaling in conjunction with the differential form of mass continuity, linear radial momentum, and boundary layer equations were used to capture the scales of the liquid film thickness and the hydrodynamic/thermal boundary layer thickness.  
Liquid film thickness, Nusselt number, and average thin-film temperature scalings and correlations were also derived. The analysis were carried out using pure scaling arguments without assuming any velocity or temperature distributions. The evaporative effects found were weak primarily as can be observed from equations having proper fractional indices for $E(r)$ (primarily the indices for $E(r)$ was $1/3$ and $1/9$ in most cases). However, successive necessary corrections have to be taken into account for larger radial coordinates where evaporative effects are significant. The average liquid film temperature scale was then calculated using the thermal boundary layer profile and the integral form of the thermal energy equation. We compared the deduced scales with experimental data available in the literature. The experimental trends of liquid film thickness and Nusselt number on different non-dimensional numbers were in harmony with the scaling laws discovered.

\phantomsection

\section*{Acknowledgments} 
The authors are thankful for the funding received from the Defence Research and Development Organization (DRDO) Chair Professorship.  




\phantomsection
\bibliographystyle{unsrt}
\bibliography{article_3.bib}


\end{document}